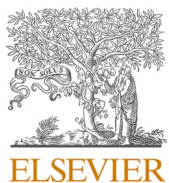
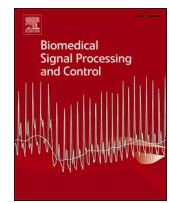

# Systolic blood pressure estimation using ECG and PPG in patients undergoing surgery

Shaoxiong Sun [a,b,c,*], Erik Bresch [a], Jens Muehlsteff [a], Lars Schmitt [a], Xi Long [a,b,*], Rick Bezemer [b,d], Igor Paulussen [a], Gerrit J. Noordergraaf [e], Ronald M. Aarts [b]

[a] *Philips Research, Eindhoven, Netherlands*
[b] *The Department of Electrical Engineering, Eindhoven University of Technology, Eindhoven, Netherlands*
[c] *The Department of Biostatistics and Health Informatics, Institute of Psychiatry, Psychology and Neuroscience, King's College London, London, UK*
[d] *Image Guided Therapy Solutions, Philips Benelux, Eindhoven, Netherlands*
[e] *Elisabeth-Tweesteden Hospital, Tilburg, the Netherlands*

ABSTRACT

*Background and Objectives:* In a significant portion of surgeries, blood pressure (BP) is often measured non-invasively in an intermittent manner. This practice has a risk of missing clinically relevant BP changes between two adjacent intermittent BP measurements. This study proposes a method to non-invasively estimate systolic blood pressure (SBP) with high accuracy in patients undergoing surgery.
*Methods:* Continuous arterial BP, electrocardiography (ECG), and photoplethysmography (PPG) signals were acquired from 29 patients undergoing surgery. After extracting 9 features from the PPG and ECG signals, we dynamically selected features upon each intermittent measurement (every 10 min) of SBP based on feature robustness and the principle of correlation-based feature selection. Finally, multiple linear regression models were built to combine these features to estimate SBP every 30 s.
*Results:* Compared to the reference SBP, the proposed method achieved a mean of difference at 0.08 mmHg, a standard deviation of difference at 7.97 mmHg, and a correlation coefficient at 0.89 (p < 0.001).
*Conclusions:* This study demonstrates the feasibility of non-invasively estimating SBP every 30 s with high accuracy during surgery by using ECG, PPG, and intermittent SBP measurements every 10 min, which meets the standard of the Association for the Advancement of Medical Instrumentation. The proposed method has the potential to enhance BP monitoring in the operating room, improving patient outcomes and experiences.

## 1. Introduction

Hemodynamic monitoring is of great importance for patients in the operating theatre. The monitoring of blood pressure (BP) is routinely implemented by either cuff-based intermittent measurement or invasive continuous measurement with a catheter. In a significant portion of operations, BP is intermittently measured with a brachial cuff. Despite the advantage of convenience and non-invasiveness, these intermittent measurements have the risk of missing clinically-relevant BP changes [1]. Thus, this method is not preferred in major surgery with the risk of rapid changes in hemodynamics.

However, catheterization has been associated with the risk of adverse effects including distal ischemia, bleeding, thrombosis, and infection, which can result in increased morbidity and costs [2–4]. These drawbacks of existing BP monitoring methods have prompted researchers to pursue non-invasive continuous solutions of arterial blood pressure (ABP) monitoring. Besides technologies such as the vascular unloading method or tonometry that both require specialized and additional equipment [5,6], photoplethysmography (PPG) has emerged as a candidate technology for this pursuit as well. By using PPG-derived features, systolic blood pressure (SBP) and/or diastolic blood pressure (DBP) can be estimated via regression models [7–14].

The most widely-studied feature for SBP estimation is pulse arrival time (PAT) [7,9,15–17]. It is defined as the time delay between an R peak in the electrocardiography (ECG) signal and a fiducial point in the PPG signal. PAT is often inversely related to the pulse wave velocity – the velocity a pulse propagates along vessels. Increased BP induces decreased elasticity of arteries, leading to an increased pulse wave velocity and therefore decreased PAT. In addition to PAT, other PPG-derived features have also been investigated, including those extracted from morphology, the first derivative, second derivative, and spectrum [7,10,18]. Although the associated physiological basis is not fully understood, several features have shown significant correlations with BP [7,19–21].

Despite the existing body of research on PPG-based methods, the achieved performances still often warrant improvements in the clinical context to meet the Association for the Advancement of Medical Instrumentation (AAMI) standard (mean difference (MD) <= 5 mmHg and standard deviation of difference (SDD) <= 8 mmHg) [22]. Liu et al. reported 8.54 (MD) ± 10.9 (SDD) for SBP estimation using PPG-derived features and support vector machine (SVM) on the MIMIC dataset collected from patients in intensive care units [23]. Another study by Zhang et al. showed 11.6 ± 8.2 mmHg for SBP estimation using PPG-derived features and SVM on the University of Queensland Vital Signs dataset collected from patients undergoing surgery [24]. Ruiz-Rodríguez et al. detailed a study where they achieved a performance of 2.98 ± 19.35 mmHg on patients in the critical care department and post-anesthesia care unit using deep learning technology [25]. Xing et al. reported 0.0 ± 11.2 mmHg with a calibration procedure for older participants using PPG-derived features and a random forest model on a dataset collected from a local community cohort in a resting state and in-hospital patients under normal medical procedures [26]. Sun et al. proposed an SBP estimation model with a performance of 0.43 ± 13.52 mmHg for young healthy subjects undertaking physical exercise with minimal model initialization and multiple linear regression [7]. The method described by Ding et al. achieved −0.37 ± 5.21 mmHg for healthy participants in a resting state with Finger cuff-derived Finapres blood pressure measurements as reference [15]. The performance of this method remains unknown for patients undergoing surgery where the BP changes drastically and when compared against invasive gold standard measurements.

Several works have investigated the confounding factors that may explain sub-optimal performance in these studies. To begin with, PAT, consisting of pulse transit time (PTT) and pre-ejection period, is not a reliable surrogate for pulse transit time which is associated with BP under the elastic tube assumption [16,27]. Furthermore, altered vascular smooth muscle tone, often assumed to be a constant in existing works, has a nonnegligible influence on the BP–PTT relationship [28]. Finally, the performance of calibrated BP estimation models decays over time as a result of changes in cardiovascular properties [29].

It is therefore interesting and relevant to investigate how to enhance the estimation model by utilizing previous intermittent BP measurements, often available in the operating room, and PPG-derived features that are designed and extracted to reflect cardiovascular properties. This will provide BP estimation at shorter time intervals in addition to existing cuff-based measurements at fixed longer time intervals, thereby potentially improving BP monitoring. In particular, in the present anesthetic monitors, SBP alarms often reply on pre-defined threshold values; changes in SBP are not monitored directly, neither are they used as part of an inbuilt alarm system [30]. In this work, we propose an SBP estimation method for patients undergoing surgery using ECG- and PPG-derived features and the information from previous intermittent measurements. In this method, dynamic feature selection was employed based on feature robustness and the principle of correlation-based feature selection (CFS).

## 2. Materials and methods

### 2.1. Patients

The study was reviewed and approved by the regional medical ethics committee (METC Brabant, The Netherlands, NL48421.028.14-P1409). With written informed consent, a heterogeneous group of 29 patients scheduled for major surgery was enrolled [31,32]. Characteristics of patients are shown in Table 1.

### 2.2. Protocol

Anesthesia was induced by propofol (2 mg/kg), sufentanil (0.5mcg/kg), and rocuronium (0.6 mg/kg), and maintained by means of continuous infusion of propofol and sufentanil. The depth of anesthesia was assessed using bi-spectral index (BSI) with a target range of 40–55. The patients were ventilated in a volume-controlled, pressure-limited mode with tidal volume of 6–10 ml/kg at a frequency of 10–14/minute and adjusted to maintain normocapnea. The positive end-expiratory pressure was set at 6 cm $H_2O$ and adjusted as needed. Fluid management was at the discretion of the physician. During surgery, three signals were collected: invasive ABP signals (Philips Heartstart MRx monitor) by a radial arterial catheter, the finger PPG signal obtained at the right index finger (Philips M1191B), and the ECG signal (Philips Heartstart MRx monitor).

### 2.3. Feature extraction

We extracted nine features that have previously been proven to have significant relationship with BP [7,12,17–19]. These features include PAT, the mean and variance of the first derivative in the systolic phase, PPG amplitude, pulse delay (PD), and four features defined in the second derivative. They were categorized into four groups when selecting features, as can be seen in Table 2 and Fig. 1. The criteria were based on a combination of physiological interpretation of these features and the robustness when extracting them. Specifically, we considered the order of derivatives in which fiducial points were detected for extracting the features, as higher order derivatives are more prone to noise. The first group comprises PAT, which has been extensively studied and of which its physiological association with SBP has been established [17]. PAT was defined as the time delay from an R peak in the ECG signal to a fiducial point (e.g., foot) in the same heartbeat of the PPG signal. It is, by far, among the most relevant and predictive features for SBP estimation. The second group consists of features based on detected fiducial points in the first derivative and comprises PPG amplitude and the mean and variance of the first derivative in the systolic phase. The latter two features were determined after the pulse was normalized. They were found to contribute significantly to SBP estimation [7]. The third group includes PD, which has been associated with the time delay between the forward and reflected wave. This delay has been shown to be altered by BP [33]. PD has been also found to be related to arterial stiffness by Millasseau et al. [19]. PD was defined as the time delay between the first and second peak in a PPG pulse. To detect a second peak, it is necessary to detect the zero-crossing point in the first derivative. However, the patients in this study were elderly and the second peak and dicrotic notch were often missing. In this case, a surrogate was used from the zero-crossing point in the second derivative [9]. The fourth group was created for the features that are based on the detected fiducial points in the third derivative. These features include c/a, e/a, norm a, and norm b [12,18]. The waves a, c and e are the first three peaks in the second derivative of the PPG pulse, while the waves b and e are the first two troughs in the same pulse. The ratios c/a and e/a were defined as ratio between the amplitude of wave c and wave a, and the ratio between the amplitude of wave e and wave a, respectively. Norm a and norm b were defined as the amplitude of wave a and b after the pulse was normalized, respectively. These wave ratios have been linked to arterial stiffness [12].

**Table 1**
Patient characteristics (n = 29).

| | |
|---|---|
| Age [yr] | 70.0 ± 8.9 |
| Gender (male/female) | 23/6 |
| BMI [kg/m$^2$] | 27.8 ± 9.7 |
| Height [cm] | 172.3 ± 13.7 |
| Length of operation [hours] | 4.4 ± 1.4 |
| Surgical procedures | |
| Urology: | |
|   Bricker deviation | 14 |
|   Radical prostatectomy | 3 |
|   Cystectomy | 1 |
|   Pyeloplasty | 1 |
| Vascular surgery: | |
|   FEM-Fem bypass or crossover | 4 |
|   EVAR removal and replacement | 3 |
|   PTA Femoral Artery | 1 |
|   Recanalization Iliac artery | 1 |
|   Carotid Endarterectomy | 1 |

**Table 2**
Extracted feature and their affiliated group. PAT: pulse arrival time. sp$_{mean}$: the mean of first derivative in the systolic phase. sp$_{var}$: the variance of the first derivative. c/a: the amplitude ratio between wave c and a. e/a: the amplitude ratio between wave e and a. norm a: the amplitude of wave a after normalising the pulse. norm b: the amplitude of wave b after normalising the pulse.

| 1st group | 2nd group | 3rd group | 4th group |
|---|---|---|---|
| PAT | PPG amplitude | Pulse delay | c/a |
|  |  |  | e/a |
|  | sp$_{mean}$ |  | norm a |
|  | sp$_{var}$ |  | norm b |

*2.4. Data analysis*

Signal analysis was confined to the period of mechanical ventilation. For ABP and PPG signals, signal segments with poor signal quality or with severe cardiac arrhythmia were excluded by manual selection and a dedicated software program. This program, after identifying peaks and valleys for each pulse, computed three parameters. These were the distance between neighboring peaks, the distance between neighboring valleys and the amplitude of each pulse. For each parameter, if the difference between the present value and the extrema (maximum or minimum) in the past window of 30 s prior to that pulse was larger than the discrepancy between these maximum and minimum values, this pulse was excluded. For ECG signals, large signal segments with consistent invalid detected R peaks were removed manually. Note that an eligible segment requires three signals be of acceptable signal quality simultaneously. As a result, data from six patients were removed entirely due to poor signal quality or cardiac arrhythmia for most time. From the remaining 23 patients, 91.2 h of data was found eligible for further analysis of PPV (9.8 % data of the 23 patients was excluded due to poor signal quality or cardiac arrhythmia). Note that invalid pulses can also be excluded when deriving the feature values in the epoch of estimation, as explained in the next paragraph.

In this study, the ECG- and PPG-derived features and SBP were extracted for each pulse (i.e., heartbeat). The epoch length for estimating SBP was 30 s. The extracted features and SBP were averaged using the extracted values within the interquartile range (25 % to 75 %) in the 30-second epoch. The intermittent measurement of every 10 min was simulated and derived using the extracted SBP in the 30-second epoch. As our focus is to evaluate the estimation algorithm, we therefore chose to simulate intermittent measurement using the average of beat-to-beat SBP rather than using the cuff-based measurement to avoid introducing additional errors that arise from the difference between cuff-based and catheter-based measurements. We chose the interval between intermittent measurements to be 10 min, as this is one option of the clinical practice.

The performance of the algorithm was evaluated by assessing agreement (Bland-Altman analysis), calculating pooled and patient-level root mean square error (RMSE) and Pearson's correlation coefficients in comparison with the reference SBP every 30 s.

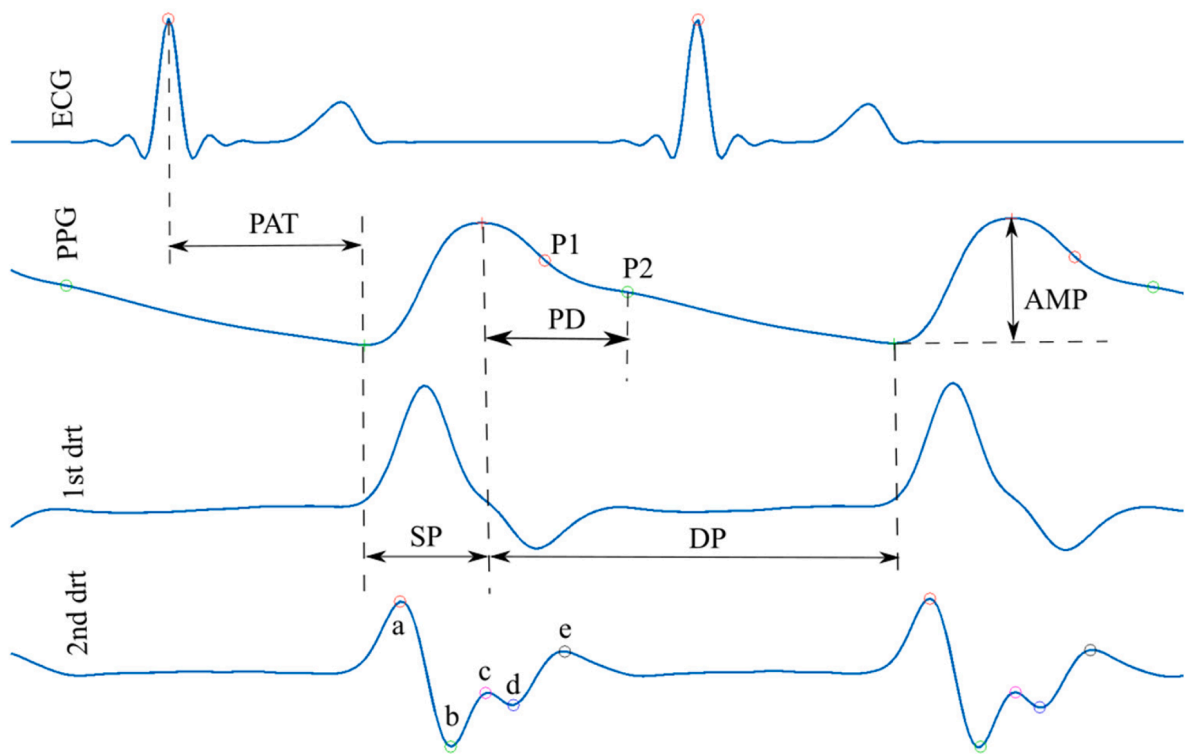

**Fig. 1.** Illustration of all features used in this study. The first panel (from top to bottom): ECG signal. The second panel: PPG signal. The third panel: the first derivative of the PPG signal. The fourth panel: the second derivative of the PPG signal. PAT: pulse arrival time. PD: pulse delay. AMP: PPG amplitude. SP: systolic phase. DP: diastolic phase. *P*1: The local maximum in the first derivative. *P*2: the local minimum in the first derivative. The waves a, c and e are the first three peaks in the second derivative of the PPG pulse, while the waves b and e are the first two troughs in that pulse.

## 2.5. Feature selection and regression model

In this work, we propose a method that includes a dynamic feature selection and regression model. To quickly initialise and apply the estimation model, the SBP estimation model was initialised by feeding the first 9 extracted PAT and SBP (every 30 s) into the regression model. A surrogate of PD was used when studying the impact of excluding ECG sensors, which will be discussed in Section 2.6. After the initialization, we selected features at the time of each intermittent measurement (every 10 min). With the selected features, we trained and applied a regression model to combine the information of these features. This model was used to estimate the SBP every 30 s for the next 10 min before the next intermittent measurement was derived. After obtaining the next intermittent SBP measurement, both the selected features and coefficients of the regression model were updated.

The feature selection scheme for the first feature is given in Fig. 2. The procedure was implemented as follows. The most recent 9 intermittent measurements (around 1.5 h) and corresponding feature values were stored. Firstly, the absolute value of correlation coefficient (ACC) between PAT and SBP was computed based on these 9 pairs of data. If the ACC was higher than 0.7 (a common threshold indicating strong correlation), then PAT was chosen as the first feature. Otherwise, the correlation between SBP and the second group (features involving the detection of the fiducial point in the first derivative) was assessed. If the maximum ACC between these features and SBP was higher than 0.7, we then performed a refined feature selection for the features with ACC higher than 0.7 (see Fig. 3.). This refined feature selection was implemented by calculating the mean correlation coefficients after bootstrapping (1000 experiments with 10 samples per experiment) for each feature. The feature with highest mean ACC was selected. This was to prevent the spurious high correlation caused by several outliers, which did not indicate its predictive power for SBP estimation. In case that no features were found to have an ACC higher than 0.7 in the first and second group, the third and fourth groups were considered. The feature selection procedure for the third and fourth group was the same as the second category. When no feature had an ACC higher than 0.7 in any of the category, we used the most recent SBP measurement as the estimation of SBP for the next 10 min.

After selecting the first feature, other features were included in the best feature subset (the features used to build the regression model) if they can provide complementary information to the first selected feature. This was realized by using correlation-based feature selection algorithm [34]. The idea is to find the best subset of features that have maximal correlation with the target variables while minimising correlation between features to reduce redundancy. The procedure is illustrated in Fig. 4. The merit indicating the strength of predicative power for a certain feature subset is given by.

$$merit = \frac{r_{sf_1} + \cdots + r_{sf_k}}{\sqrt{k + 2*(r_{f_1f_2} + \cdots + r_{f_if_j} + \cdots + r_{f_kf_1})}} \quad (1)$$

where $r_{sf_i}$ denotes ACC between *i*-th features and SBP, $r_{f_if_j}$ denotes ACC between *i*-th and *j*-th features, and *k* is the number of features.

The new merits were calculated for the features already in the best feature subset combined with each newly added feature and the highest merit among all calculated merits was identified. This highest merit was then compared with the stored best merit (the merit for the current best subset). If the current merit was larger than the stored one, the current feature was added to the best feature subset and the stored merit was overwritten by the new highest merit. Because we only used 9 points to build the regression model, to avoid overfitting, we set the maximum number of features to be 3, after comparing this with maximum number of features being 1 and 2 using this feature selection scheme, detailed in Section 3. This choice of 3 was also suggested by [35], where the square root of the number of samples was preferred when features were correlated.

After identifying the best feature subset, a multiple linear regression model was built by using SBP and features from the most recent 9 intermittent measurements. While an insufficient number of recent intermittent measurements provide inadequate information for the regression model, an excessive number of recent intermittent measurements make the model unnecessarily depend on the past, losing flexibility and strength in the dynamic feature selection. The choice of 9 was made based on the model performance, as detailed in Section 3. The regression coefficients for each feature were determined collectively in this linear fitting. After acquiring the coefficients for each feature, an offset was added to match the most recent intermittent measurement. This was to ensure that the model fitted perfectly with the most recent measurement.

## 2.6. Performance comparisons with other methods

To further evaluate the performance of the proposed method, we first compared it with that of two existing methods. These two methods also provided SBP estimation every 30 s. In the first method, SBP was estimated using the most recent intermittent SBP measurement until a new SBP intermittent measurement was acquired. This is essentially the zero-order hold of the most recent measurement, which is in line with what current clinical practice implies where the clinicians can only rely on most recent measurement of SBP when making clinical decisions. Thus, the first method was termed zero-order hold. In the second method, the model was built using only PAT as the feature, termed PAT-only. Similar to the proposed methods, the coefficients of regression model were estimated using these 9 intermittent measurements and updated every 10 min upon a new intermittent measurement coming in. To assess the utility of ECG sensors, we also built an estimation model using the proposed dynamic feature selection scheme but based on PPG-derived features only, termed PPG-only. To demonstrate the usefulness of the proposed dynamic feature selection scheme, we implemented a method based on principal component analysis (PCA). Similar to the proposed method, we allowed flexibility to select the exact number of the included components in the regression model. The number of components was determined using explained variance ratio. We chose 70 %, in line with the recommendation in [36]. As described in Section 2.5., we also compared the proposed method with the ones with the proposed dynamic feature selection scheme but with the conditions of 1 or 2 features at maximum. Finally, to demonstrate the necessity and improvement of the proposed dynamic model updating scheme, we compared the 5 SBP estimations right after updating and the 5 SBP estimations closest to the time of the next updating.

## 2.7. Statistical analysis

To implement pairwise statistical comparisons between the proposed method with the other methods under consideration, we applied Wilcoxon signed rank test on the paired RMSE and correlation coefficients at the patient level. The other methods included the zero-order hold and PAT-only, PPG-only, the proposed dynamic feature selection scheme with the maximum number of features being set to be 1 and 2, the PCA-based model. To avoid type II errors due to the small sample size of this study, we did not perform multiple testing corrections. A $p < 0.05$ was deemed statistically significant.

## 3. Results

The SBP estimation performance of the methods in comparisons can be found in Table 3. The proposed method outperformed all the other methods. The pooled RMSE was 7.97 mmHg and correlation coefficient 0.89 ($p < 0.001$). The proposed method achieved a mean and SD of difference was 0.08 and 7.97 respectively, meeting the AAMI standard.

When compared to the zero-order hold and PAT-only methods, the proposed method showed statistically significant difference in both RMSE and correlation coefficients. The PPG-only method performed worse than the proposed method where both ECG and PPG were required. The differences in the RMSE and correlation coefficients were both statistically significant. The boxplots showing the comparisons between the proposed method and these three methods are given in Fig. 5. While the proposed method showed the lowest median and an interquartile range smaller than the zero-order hold and PPG-only method in the RMSE, the proposed method showed the highest median and an interquartile range smaller than the zero-order hold and PAT-only method in the correlation coefficients. Fig. 6. shows the Bland-Altman plot between the estimated and measured SBP for zero-order hold, PAT-only, PPG-only and the proposed method. The proposed method showed smaller limits of agreements compared to the other three.

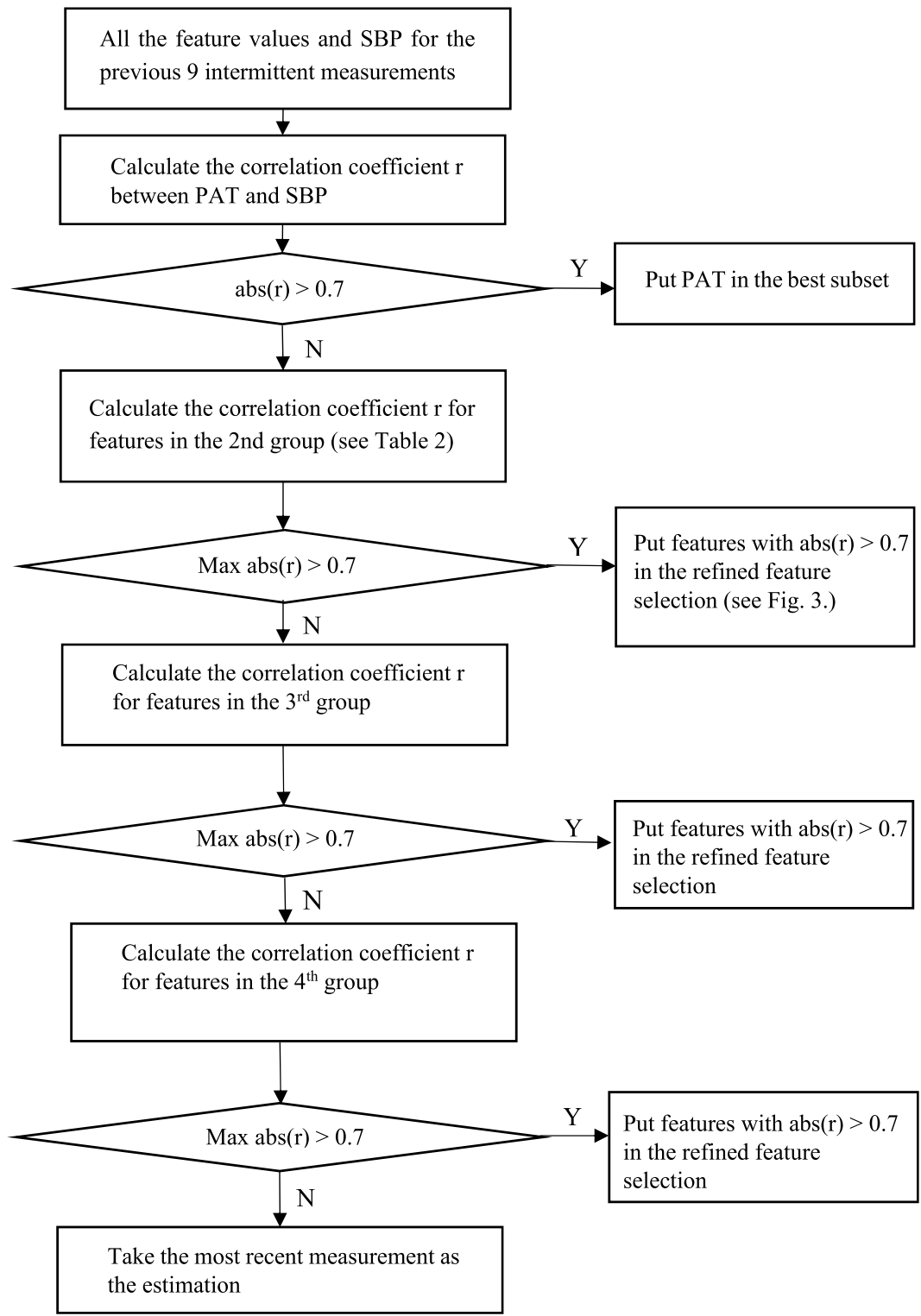

**Fig. 2.** The flowchart for selecting the first feature in the proposed method. The absolute value of correlation coefficient (ACC) between PAT and SBP was computed based on the 9 pairs of most recent data. If the ACC was higher than 0.7, PAT was chosen as the first feature. Otherwise, the correlation between SBP and the second group was assessed. If the maximum ACC between these features and SBP was higher than 0.7, we then performed a refined feature selection for the features with ACC higher than 0.7 (see Fig. 3.). In case that no features were found to have an ACC higher than 0.7 in the first and second group, the third and fourth groups were considered. The feature selection procedure for the third and fourth group was the same as the second category. When no feature had an ACC higher than 0.7 in any of the category, the most recent SBP measurement was used as the estimation of SBP for the next 10 min.

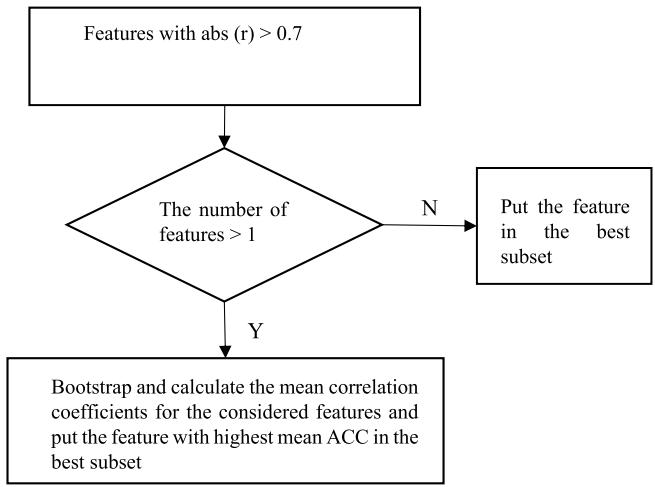

**Fig. 3.** The refined feature selection for selecting the first feature in Fig. 2. The mean correlation coefficients were calculated after bootstrapping (1000 experiments with 10 samples per experiment) for each feature. The feature with highest mean absolute correlation coefficient was selected.

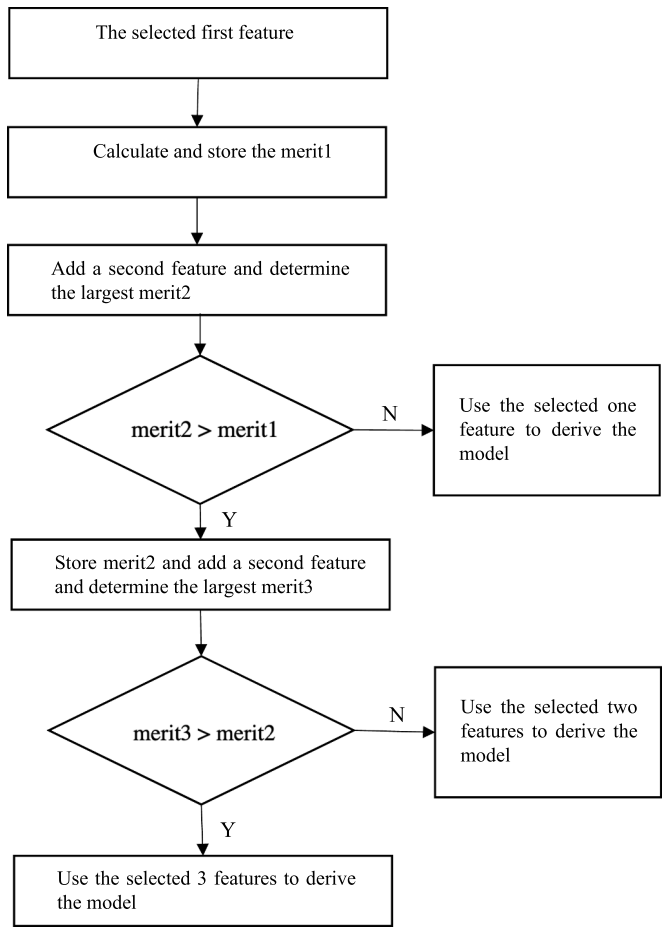

**Fig. 4.** The selection strategy for the second and third features. Merit1 was calculated with the first selected feature. A second feature from the rest of extracted features was added and combined with the already selected first feature. The best merit2 from all combination was calculated based on the principal of correlation-based feature selection (see Eq. (1)). If merit2 > merit1, the corresponding feature was selected. The Third feature was selected in the same manner.

When it comes to the impact of the maximum number of features included in the proposed dynamic feature selection scheme, the proposed method (3 features at maximum) had better performance compared to that with 1 or 2 features. The difference in correlation coefficients was found to be statistically significant, while that in RMSE showed p-values of 0.06 and 0.09, respectively. The proposed method also outperformed PCA-based method showing statistically significant difference in both correlation coefficients and RMSE. Finally, the RMSE for the 5 SBP estimations right after updating the model was 5.42 mmHg, while the RMSE for 5 SBP estimations closest to the time of next updating was 9.38 mmHg, demonstrating the necessity and improvements of online model updating.

Fig. 7. shows an example on performance comparisons between the zero-order hold, PAT-only, and the proposed methods. It is shown that the proposed method approached better the value and trend of measured SBP compared to the other two methods. Fig. 8. presents the impact on the RMSE and correlation coefficients of the number of recent intermittent measurements included when building the model. The RMSE decreased with the number of recent intermittent measurements, reached minimum at 9, and then increased with the number of the measurements. Similarly, the correlation coefficients reached its maximum at 9. The statistics of the selected features for the proposed method is presented in Table 4. It is shown that three features were selected in a majority of cases and PAT was often chosen as one of these three features. Next to best feature subsets comprising three features, the best feature subset comprising one feature was chosen to build the regression model, where PAT was the most selected feature. In very few cases, two features were selected to build the regression model.

## 4. Discussion

In this study, we propose a novel method utilizing previous intermittent SBP measurements and ECG- and PPG-derived features to build regression models for continuous SBP estimation (every 30 s) for patients undergoing surgery in the operating room. Based on the physiological understanding of these features and robustness when extracting them, we selected the first feature (the most important feature). Next, we selected the features that provided complementary information based on the principle of CFS. The results show that the proposed method outperformed the zero-order hold, PAT-only, PPG-only and PCA-based methods. Furthermore, it achieved a high accuracy meeting the AAMI standard [22]. The proposed method will provide continuous SBP estimation at short time interval between the intermittent BP measurements in surgeries, thereby potentially improving BP monitoring.

In a significant portion of operations, invasive beat-to-beat BP monitoring is not necessary, and BP is only intermittently measured using a brachial cuff. Although non-invasive and convenient, this may hinder clinicians from observing clinically relevant changes in the BP in the interval between intermittent BP measurements. It is therefore important and valuable to also monitor BP between adjacent intermittent measurements. In this study, we achieved this goal by building an ECG- and PPG-based regression model that can provide SBP estimation every 30 s. Furthermore, we demonstrated that the proposed method yielded estimation that was in a good agreement and well correlated with the reference SBP, aided by the calibration and recalibration of the model using the intermittent measurements simulated using the continuous BP. The SD of difference was within the acceptable threshold of 8 mmHg according to the AAMI standard [22]. The achieved performance was also superior compared to existing works [7,23,24] in the mean and SD of difference and compared to the work by Xing and et al. [26] in the SD of difference.

**Table 3**
The SBP estimation performance for all estimation models under consideration.

|  | Pooled RMSE (mmHg) | Bland-Altman analysis (Bias ± SD) [mmHg] | Pooled correlation Coefficients | Median difference in RMSE [mmHg] | Median difference in correlation coefficients |
| --- | --- | --- | --- | --- | --- |
| Zero-order hold | 13.12 | −0.07 ± 13.12 | 0.71*** | −4.32*** | 0.21*** |
| PAT-only | 10.18 | 0.28 ± 10.18 | 0.82*** | −1.11** | 0.05*** |
| PPG-only | 9.92 | 0.38 ± 9.91 | 0.84*** | −1.46** | 0.04** |
| **Proposed method (DFS[a] with 3 features at maximum)** | **7.97** | **0.08 ± 7.97** | **0.89*** | – | – |
| DFS[a] scheme with 2 features at maximum | 8.30 | 0.09 ± 8.30 | 0.88*** | −0.42 (0.09) | 0.01* |
| DFS[a] scheme with 1 feature | 8.64 | 0.17 ± 8.64 | 0.87*** | −0.54 (0.06) | 0.02* |
| PCA | 9.92 | 0.23 ± 9.92 | 0.83*** | −1.00** | 0.05*** |

\* statistically significance ($p < 0.05$).
\*\* statistically significance ($p < 0.01$).
\*\*\* statistically significance ($p < 0.001$).
[a] The proposed dynamic feature selection scheme.

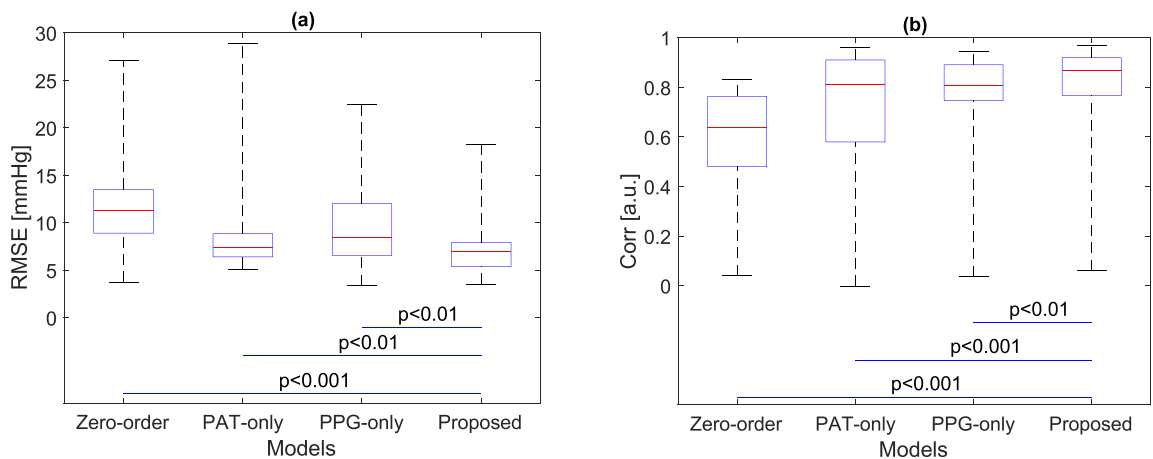

**Fig. 5.** Comparisons of model performance between different models (a) Comparisons in RMSE (b) Comparisons in correlation coefficients. The p-value from signed-rank test is presented on the solid line. Horizontal lines in the boxes: median values. Edges of boxes: 25th and 75th percentiles. Whiskers: maximum and minimum.

Estimating and controlling SBP are of clinical importance, as elevated SBP has been associated with cardiovascular risk and increased left ventricular mass and hypertrophy [37–39]. The SBP estimation models during various conditions such as during surgery as described in this work, during physical exercise [7], and during resting periods [15] that can potentially monitor SBP for extended periods, coupled with cost-efficient identifying left ventricular structure and function using image segmentation techniques [40–44], may lead to a deeper understanding of relationship between BP and cardiovascular systems.

A few features have been applied to estimate SBP in the existing literature. The extraction of some of these features is more sensitive to noise when detecting fiducial points has to be done in higher derivatives. This is because the relevant signal is often attenuated in the higher order derivative, while noise is often largely present. In this study, we designed a method where features were categorized based on the used derivatives when extracting these features. After selecting the best feature based on this consideration, additional features were chosen based on the idea that a relevant feature should correlate reasonably well with SBP while not introducing redundancy in the presence of existing features. The results show that PAT is most likely to be selected. This is in accordance with literature, as PAT is often shown to be correlated well with BP and has a physiological basis. In addition to PAT, other features such as $sp_{var}$ and e/a in the second derivative are often chosen, which is in line with the finding in our previous work [7]. We also showed the model using features derived only from PPG showed a reduced performance, further highlighting the contribution of PAT. More recently, there have been works on using ECG alone to build BP estimation models [45,46]. While this is beyond the scope of this work, future work might also study the performance of models including features extracted from ECG alone.

While computational complexity has been rarely reported in existing studies, the application of advanced machine learning or deep learning models can result in high computational complexity. In this study, as the proposed method included dynamic feature selection and a linear regression with 9 data points at each time of recalibration, the computational complexity of the proposed method is O(N), lower than existing studies using SVM and deep learning models [23–25].

To expedite the model initialization process, the model was initialised by taking the first 9 measurements of SBP and calculated features every 30 s. When this is challenging with a typical brachial cuff, an immediately calibrated finger cuff based SBP measurement can be employed temporarily as a surrogate for this short period of time. In the case of this finger cuff introducing potential bias, it will not accumulate and propagate to degrade future model performance at later times, as the model always takes the most recent 9 measurements for model training after initialisation.

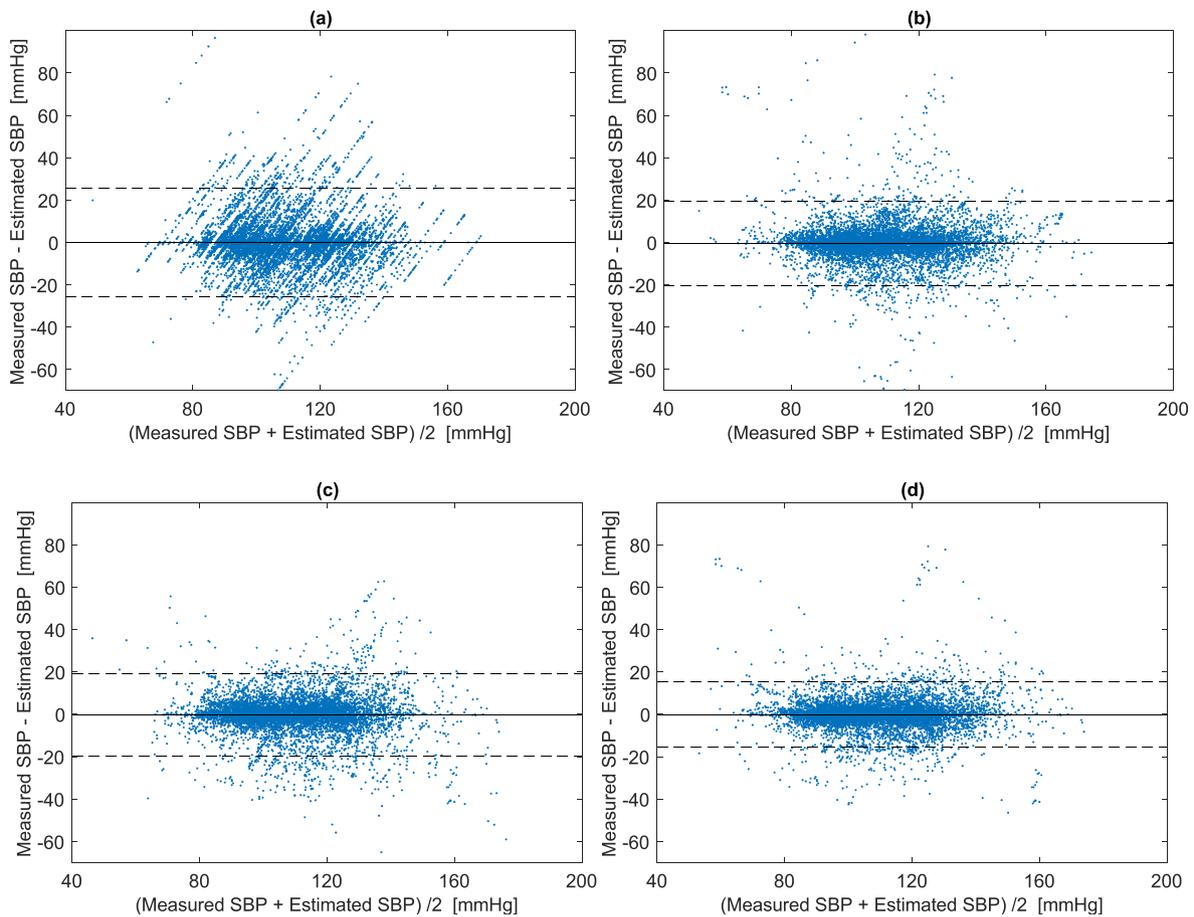

**Fig. 6.** Bland-Altman plot of the measured SBP and estimated SBP using the proposed method. (a) Zero-order hold. (b) PAT-only. (c) PPG-only method. (d) Proposed method. The solid line corresponds to the bias (mean difference) and the dotted lines correspond to the limits of agreement (1.96 × standard deviation of difference).

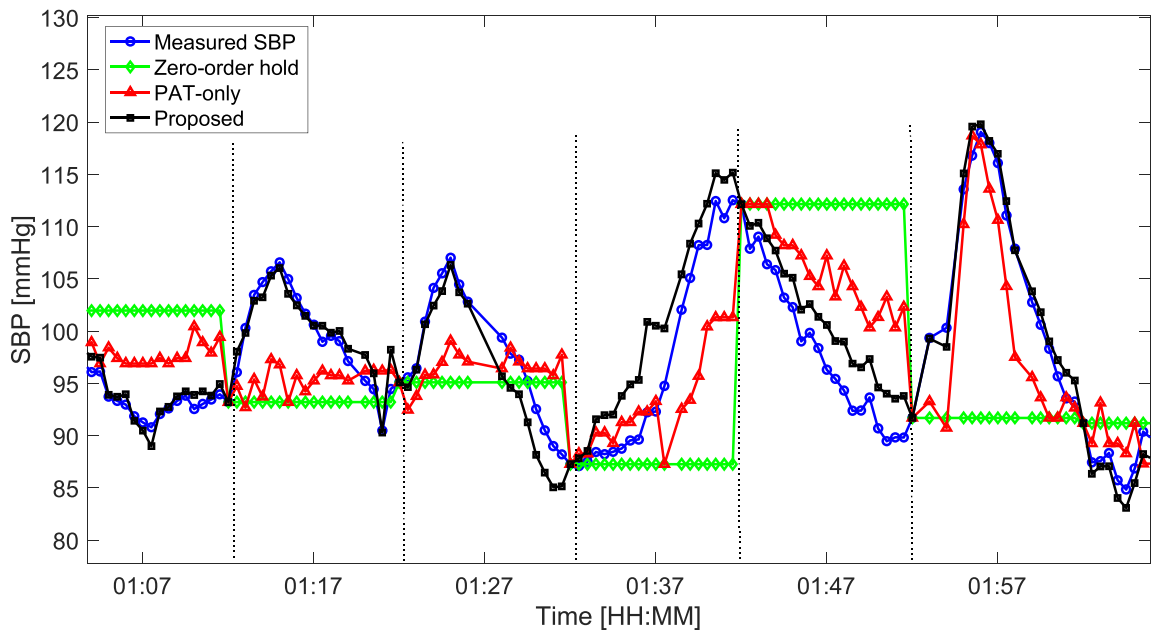

**Fig. 7.** An example of estimated SBP in comparison to measured SBP. The methods in comparison are estimation zero-order hold, PAT-only, and the proposed method. The dashed lines marked the time of recalibration.

Our study has several limitations. Due to the small size of the sample, we were unable to apply other regression models involving hyper-parameter tuning (e.g., LASSO). With the future acquisition of a larger dataset, these parameters can be optimized and an improvement in the estimation performance might be expected. Second, the intermittent measurement in this study was simulated and derived using the continuous BP measurement instead of directly using cuff-based measurements. This was to avoid introducing errors when comparing estimated and measured SBP and to facilitate the validation of the proposed method. Future work will focus on directly using intermittent cuff-based BP measurements to calibrate and recalibrate the estimation model and compare the performances. Third, in other clinical scenarios with non-sedated patients, a fixed measurement-interval of 10 min might be uncomfortable for some patients. In those scenarios, future research might be dedicated to devising smart triggering strategy where the BP measurement is only triggered when the estimation algorithm is likely to fail. By doing this, we can maximize the interval between each measurement. Fourth, we excluded data from 6 patients due to poor signal quality or cardiac arrhythmia to avoid confounding factors to the algorithm design. Future work might focus on how to reliably extract features in these conditions, in particular extracting PAT in the presence of abnormal heart's electrical activities. Finally, this work focuses on SBP estimation for its high clinical relevance in anesthetic monitoring. Future work might also extend this approach to SBP and DBP monitoring in different settings.

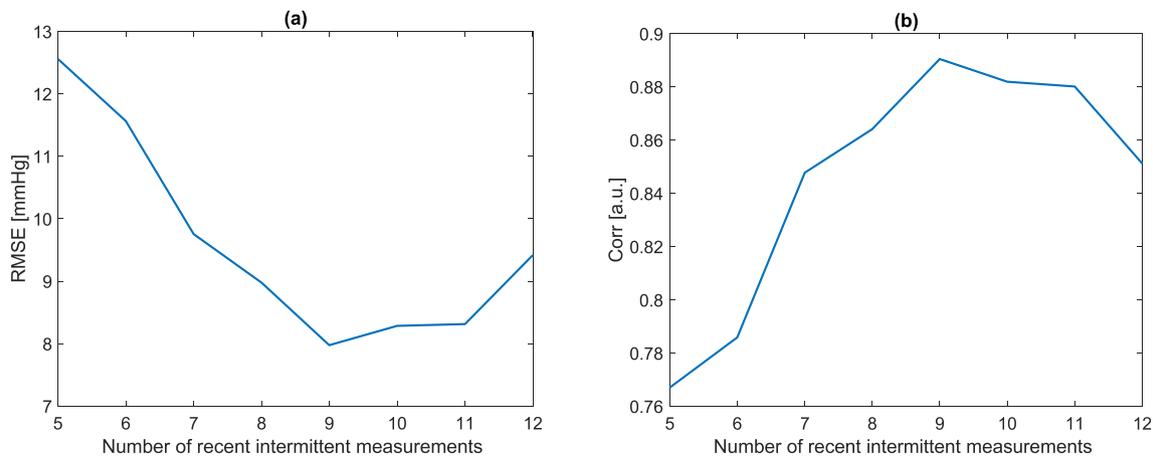

**Fig. 8.** The impact of the number recent intermittent measurements included when building the model on the model performance. (a) The impact on root mean square errors. (b) The impact on correlation coefficients.

**Table 4**
The statistics of the selected features for the proposed method.

|  | Occurrence frequency | Top 3 occurrence |
|---|---|---|
| One feature | 19.1 % | PAT<br>PPG amplitude<br>$sp_{var}$ |
| Two features | 1.5 % | e/a, PAT<br>$sp_{var}$, PAT<br>PPG amplitude, PAT |
| Three features | 79.4 % | $sp_{mean}$, norm b, PAT<br>c/a, e/a, PAT<br>$sp_{var}$, c/a, PAT |

## 5. Conclusion

In summary, this study presents a novel method that can provide SBP estimation every 30 s based on recent intermittent measurements of SBP with a high accuracy meeting the AAMI standard. This was achieved by considering and modeling the changes in cardiovascular properties using ECG- and PPG-derived features and applying a dynamic feature selection scheme based on feature robustness and the principle of CFS. The proposed method outperformed the zero-order hold, PAT-only, PPG-only and PCA-based methods. The proposed method also showed superior performance compared to results reported in existing literature. The proposed method has the potential to enhance BP monitoring in the operating room, improving patient outcomes and experiences.

## CRediT authorship contribution statement

**Shaoxiong Sun:** Conceptualization, Methodology, Software, Formal analysis, Investigation, Writing – original draft, Writing – review & editing. **Erik Bresch:** Investigation, Writing – review & editing. **Jens Muehlsteff:** Investigation, Writing – review & editing. **Lars Schmitt:** Investigation, Writing – review & editing. **Xi Long:** Writing – review & editing. **Rick Bezemer:** Supervision, Writing – review & editing. **Igor Paulussen:** Investigation. **Gerrit J. Noordergraaf:** Investigation, Writing – review & editing. **Ronald M. Aarts:** Supervision, Writing – review & editing.

## Declaration of Competing Interest

The authors declare that they have no known competing financial interests or personal relationships that could have appeared to influence the work reported in this paper [Erik Bresch, Jens Muehlsteff, Lars Schmitt, Xi Long, Igor Paulussen, and Rick Bezemer are employed by Philips.].

## Data availability

The authors do not have permission to share data.

## Acknowledgements

The authors wish to express their gratitude to Arthur Bouwman and Erik Korsten for reviewing the manuscript and to Wouter Peeters for collecting and pre-processing data.